\pdfoutput=1

\documentclass[12pt,a4paper]{article}

\usepackage{ifthen} 
\newboolean{pdflatex}
\setboolean{pdflatex}{true} 

\newboolean{articletitles}
\setboolean{articletitles}{true} 

\newboolean{uprightparticles}
\setboolean{uprightparticles}{false} 

\def\paperauthors{LHCb collaboration} 
\def\paperasciititle{Observation of a narrow pentaquark state, Pc(4312)+, and of two-peak structure of the Pc(4450)+} 
\def\papertitle{Observation of a narrow pentaquark state, $P_c(4312)^+$, and of two-peak structure of the $P_c(4450)^+$} 
\def\paperkeywords{{High Energy Physics}, {LHCb}} 
\def\papercopyright{\the\year\ CERN for the benefit of the LHCb collaboration} 
\def\paperlicence{CC-BY-4.0 licence}
\def\paperlicenceurl{https://creativecommons.org/licenses/by/4.0/}

\usepackage[top=1in, bottom=1.25in, left=1in, right=1in]{geometry}

\usepackage{microtype}
\usepackage{lineno}  
\usepackage{xspace} 
\usepackage{caption} 

\usepackage{graphicx}  
\usepackage{color}
\usepackage{colortbl}
\graphicspath{{./figs/}} 
\DeclareGraphicsExtensions{.pdf,.PDF,png,.PNG}

\usepackage{amsmath} 
\usepackage{amssymb}
\usepackage{amsfonts}
\usepackage{upgreek} 

\newcommand*\patchAmsMathEnvironmentForLineno[1]{%
\expandafter\let\csname old#1\expandafter\endcsname\csname #1\endcsname
\expandafter\let\csname oldend#1\expandafter\endcsname\csname
end#1\endcsname
 \renewenvironment{#1}%
   {\linenomath\csname old#1\endcsname}%
   {\csname oldend#1\endcsname\endlinenomath}%
}
\newcommand*\patchBothAmsMathEnvironmentsForLineno[1]{%
  \patchAmsMathEnvironmentForLineno{#1}%
  \patchAmsMathEnvironmentForLineno{#1*}%
}
\AtBeginDocument{%
\patchBothAmsMathEnvironmentsForLineno{equation}%
\patchBothAmsMathEnvironmentsForLineno{align}%
\patchBothAmsMathEnvironmentsForLineno{flalign}%
\patchBothAmsMathEnvironmentsForLineno{alignat}%
\patchBothAmsMathEnvironmentsForLineno{gather}%
\patchBothAmsMathEnvironmentsForLineno{multline}%
\patchBothAmsMathEnvironmentsForLineno{eqnarray}%
}

\usepackage{hyperxmp}

\usepackage[pdftex,
            pdfauthor={\paperauthors},
            pdftitle={\paperasciititle},
            pdfkeywords={\paperkeywords},
            pdfcopyright={Copyright (C) \papercopyright},
            pdflicenseurl={\paperlicenceurl}]{hyperref}

\usepackage[colorinlistoftodos,textsize=scriptsize]{todonotes}

\usepackage[all]{hypcap} 

\usepackage{xspace} 
\usepackage{upgreek}

\def\lhcb   {\mbox{LHCb}\xspace}

\def\MagUp {\mbox{\em Mag\kern -0.05em Up}\xspace}

\ifthenelse{\boolean{uprightparticles}}%
{

 \def\Ppsi        {\ensuremath{\uppsi}\xspace}

 \def\PDelta      {\ensuremath{\Delta}\xspace}                 
 \def\PXi         {\ensuremath{\Xi}\xspace}                 
 \def\PLambda     {\ensuremath{\Lambda}\xspace}                 
 \def\PSigma      {\ensuremath{\Sigma}\xspace}                 
 \def\POmega      {\ensuremath{\Omega}\xspace}                 
 \def\PUpsilon    {\ensuremath{\Upsilon}\xspace}

 \def\PB      {\ensuremath{\mathrm{B}}\xspace}                 
                  
 \def\PD      {\ensuremath{\mathrm{D}}\xspace}

 \def\PJ      {\ensuremath{\mathrm{J}}\xspace}                 
 \def\PK      {\ensuremath{\mathrm{K}}\xspace}

 \def\Pb      {\ensuremath{\mathrm{b}}\xspace}                 
 \def\Pc      {\ensuremath{\mathrm{c}}\xspace}

 \def\Pi      {\ensuremath{\mathrm{i}}\xspace}

 \def\thebaroffset{0.0em}
}
{

 \def\Ppsi        {\ensuremath{\psi}\xspace}                 
                  
 \mathchardef\PDelta="7101
 \mathchardef\PXi="7104
 \mathchardef\PLambda="7103
 \mathchardef\PSigma="7106
 \mathchardef\POmega="710A
 \mathchardef\PUpsilon="7107
                  
 \def\PB      {\ensuremath{B}\xspace}                 
                  
 \def\PD      {\ensuremath{D}\xspace}

 \def\PJ      {\ensuremath{J}\xspace}                 
 \def\PK      {\ensuremath{K}\xspace}

 \def\Pb      {\ensuremath{b}\xspace}                 
 \def\Pc      {\ensuremath{c}\xspace}

 \def\Pi      {\ensuremath{i}\xspace}

 \def\thebaroffset{0.18em}
}
\newcommand{\offsetoverline}[2][\thebaroffset]{\kern #1\overline{\kern -#1 #2}}%

\makeatletter
\ifcase \@ptsize \relax
  \newcommand{\miniscule}{\@setfontsize\miniscule{4}{5}}
\or
  \newcommand{\miniscule}{\@setfontsize\miniscule{5}{6}}
\or
  \newcommand{\miniscule}{\@setfontsize\miniscule{5}{6}}
\fi
\makeatother

\DeclareRobustCommand{\optbar}[1]{\shortstack{{\miniscule (\rule[.5ex]{1.25em}{.18mm})}
  \\ [-.7ex] $#1$}}

\def\cquark    {{\ensuremath{\Pc}}\xspace}

\def\bquark    {{\ensuremath{\Pb}}\xspace}

\def\kaon    {{\ensuremath{\PK}}\xspace}

\def\KorKbar {\kern \thebaroffset\optbar{\kern -\thebaroffset \PK}{}\xspace}

\def\Km      {{\ensuremath{\kaon^-}}\xspace}

\def\Dbar    {{\ensuremath{\offsetoverline{\PD}}}\xspace}

\def\Db      {{\ensuremath{\Dbar}}\xspace}
\def\DorDbar {\kern \thebaroffset\optbar{\kern -\thebaroffset \PD}\xspace}

\def\Dzb     {{\ensuremath{\Dbar{}^0}}\xspace}

\def\Dstarb  {{\ensuremath{\Dbar{}^*}}\xspace}

\def\Dstarzb {{\ensuremath{\Dbar{}^{*0}}}\xspace}

\def\BorBbar {\kern \thebaroffset\optbar{\kern -\thebaroffset \PB}\xspace}

\def\jpsi     {{\ensuremath{{\PJ\mskip -3mu/\mskip -2mu\Ppsi\mskip 2mu}}}\xspace}

\def\Y#1S{\ensuremath{\PUpsilon{(#1S)}}\xspace}

\def\Lz          {{\ensuremath{\PLambda}}\xspace}

\def\LorLbar     {\kern \thebaroffset\optbar{\kern -\thebaroffset \PLambda}\xspace}

\def\Sigmares    {{\ensuremath{\PSigma}}\xspace}

\def\Lc          {{\ensuremath{\Lz^+_\cquark}}\xspace}

\def\Lb           {{\ensuremath{\Lz^0_\bquark}}\xspace}

\def\to                 {\ensuremath{\rightarrow}\xspace}

\def\AT#1     {\ensuremath{A_{\mathrm{T}}^{#1}}\xspace}           

\def\C#1      {\ensuremath{\mathcal{C}_{#1}}\xspace}                       
\def\Cp#1     {\ensuremath{\mathcal{C}_{#1}^{'}}\xspace}                    
\def\Ceff#1   {\ensuremath{\mathcal{C}_{#1}^{\mathrm{(eff)}}}\xspace}        
\def\Cpeff#1  {\ensuremath{\mathcal{C}_{#1}^{'\mathrm{(eff)}}}\xspace}       
\def\Ope#1    {\ensuremath{\mathcal{O}_{#1}}\xspace}                       
\def\Opep#1   {\ensuremath{\mathcal{O}_{#1}^{'}}\xspace}                    

\newcommand{\aunit}[1]{\ensuremath{\text{\,#1}}}       

\newcommand{\tev}{\aunit{Te\kern -0.1em V}\xspace}
\newcommand{\gev}{\aunit{Ge\kern -0.1em V}\xspace}
\newcommand{\mev}{\aunit{Me\kern -0.1em V}\xspace}
\newcommand{\kev}{\aunit{ke\kern -0.1em V}\xspace}
\newcommand{\ev}{\aunit{e\kern -0.1em V}\xspace}
\newcommand{\mevc}{\ensuremath{\aunit{Me\kern -0.1em V\!/}c}\xspace}
\newcommand{\gevc}{\ensuremath{\aunit{Ge\kern -0.1em V\!/}c}\xspace}
\newcommand{\mevcc}{\ensuremath{\aunit{Me\kern -0.1em V\!/}c^2}\xspace}
\newcommand{\gevcc}{\ensuremath{\aunit{Ge\kern -0.1em V\!/}c^2}\xspace}

\def\fb   {\ensuremath{\aunit{fb}}\xspace}
\def\invfb   {\ensuremath{\fb^{-1}}\xspace}

\def\gsim{{~\raise.15em\hbox{$>$}\kern-.85em
          \lower.35em\hbox{$\sim$}~}\xspace}
\def\lsim{{~\raise.15em\hbox{$<$}\kern-.85em
          \lower.35em\hbox{$\sim$}~}\xspace}

\xspace

\def\tell1  {TELL1\xspace}
\def\ukl1   {UKL1\xspace}

\usepackage{cite} 
\usepackage{mciteplus}

\usepackage{longtable} 

\def\Sigmac          {{\ensuremath{\Sigmares^+_\cquark}}\xspace}

\newboolean{prl}
\setboolean{prl}{false} 

\begin{document}

\renewcommand{\thefootnote}{\fnsymbol{footnote}}
\setcounter{footnote}{1}


\begin{titlepage}
\pagenumbering{roman}

\vspace*{-1.5cm}
\centerline{\large EUROPEAN ORGANIZATION FOR NUCLEAR RESEARCH (CERN)}
\vspace*{1.5cm}
\noindent
\begin{tabular*}{\linewidth}{lc@{\extracolsep{\fill}}r@{\extracolsep{0pt}}}
\ifthenelse{\boolean{pdflatex}}
{\vspace*{-1.5cm}\mbox{\!\!\!\includegraphics[width=.14\textwidth]{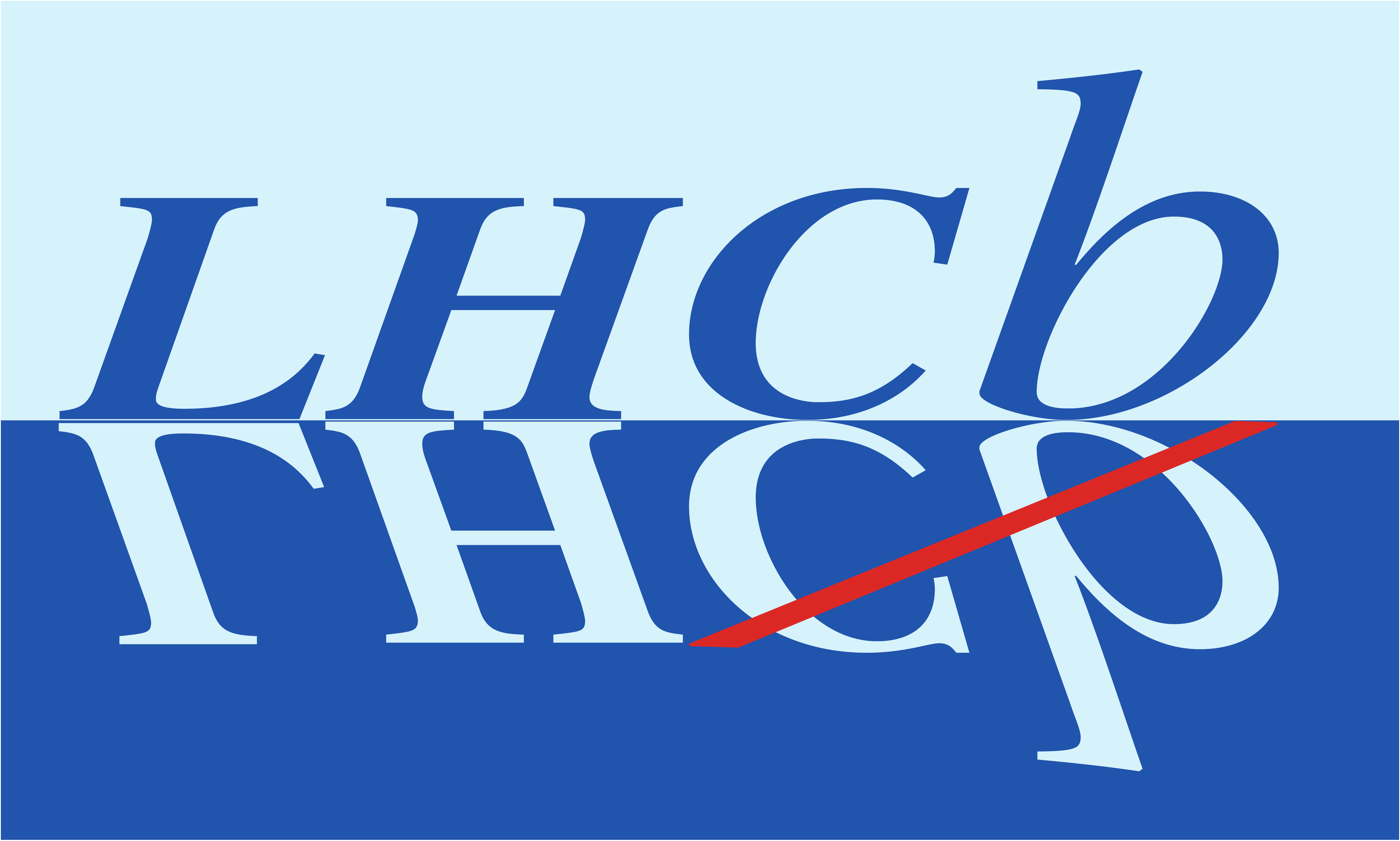}} & &}%
{\vspace*{-1.2cm}\mbox{\!\!\!\includegraphics[width=.12\textwidth]{lhcb-logo.eps}} & &}%
\\
 & & CERN-EP-2019-058 \\  
 & & LHCb-PAPER-2019-014 \\  
 & & 14 May 2019 \\ 
 & & \\
\end{tabular*}

\vspace*{4.0cm}

{\normalfont\bfseries\boldmath\huge
\begin{center}
  \papertitle 
\end{center}
}

\vspace*{2.0cm}

\begin{center}
\paperauthors\footnote{Authors are listed at the end of this paper.}
\end{center}

\vspace{\fill}

\begin{abstract}
  \noindent
   A  narrow pentaquark state, $P_c(4312)^+$, decaying to $\jpsi p$ is discovered 
 with a statistical significance of  $7.3\sigma$  
 in a data sample of ${\Lb\to\jpsi p\Km}$ decays which is an order of magnitude larger than that previously analyzed by the LHCb collaboration. 
 The $P_c(4450)^+$ pentaquark structure formerly reported by LHCb  is confirmed and observed to consist of two narrow overlapping peaks, $P_c(4440)^+$ and $P_c(4457)^+$, where the statistical 
 significance of this two-peak interpretation is  $5.4\sigma$. 
 Proximity of the $\Sigmac\Dzb$ and  $\Sigmac\Dstarzb$ thresholds to the observed narrow peaks suggests that they play an important role in the dynamics of these states.

\end{abstract}

\vspace*{2.0cm}

\begin{center}
  Accepted by Phys.\ Rev.\ Lett.
\end{center}

\vspace{\fill}

{\footnotesize 
\centerline{\copyright~\papercopyright. \href{\paperlicenceurl}{\paperlicence}.}}
\vspace*{2mm}

\end{titlepage}


\newpage
\setcounter{page}{2}
\mbox{~}
%
%
%
%

\cleardoublepage


\renewcommand{\thefootnote}{\arabic{footnote}}
\setcounter{footnote}{0}



\pagestyle{plain} 
\setcounter{page}{1}
\pagenumbering{arabic}


%

\def\mkp{m_{Kp}}
\def\mjpsip{m_{\jpsi p}}
A major turning point in exotic baryon spectroscopy was achieved at the Large Hadron Collider when, 
from an analysis of Run 1 data, the LHCb collaboration reported the observation of significant $\jpsi p$ pentaquark structures in $\Lb\to\jpsi p\Km$ decays (inclusion of charge-conjugate processes is implied throughout). 
A model-dependent six-dimensional amplitude analysis of invariant masses and decay angles describing the $\Lb$ decay revealed a $P_c(4450)^+$ structure peaking at $4449.8\pm 1.7\pm 2.5\mev$ with a width of $39\pm 5\pm 19\mev$ and a fit fraction of ($4.1\pm0.5\pm1.1)\%$~\cite{LHCb-PAPER-2015-029}.
Even though not apparent from the $m_{\jpsi p}$ distribution alone, the amplitude analysis also required a second broad $\jpsi p$ state to obtain a good description of the data, which peaks at $4380\pm 8\pm 29$\mev with a width of $205\pm 18\pm 86$\mev and a fit fraction of ($8.4\pm0.7\pm4.2$)\%.
Furthermore, the exotic hadron character of the $\jpsi p$ structure  near 4450\mev was demonstrated in a model-independent way in Ref.~\cite{LHCb-PAPER-2016-009}, where it was shown to be too narrow to be accounted for by  $\Lz^* \to p\Km$ reflections ($\Lz^*$ denotes $\Lz$ excitations).
Various interpretations of these structures have been proposed, including tightly bound $duuc\bar{c}$ pentaquark 
states \cite{Maiani:2015vwa,%
Lebed:2015tna,%
Anisovich:2015cia,%
Li:2015gta,%
Ghosh:2015ksa,%
Wang:2015epa,
Zhu:2015bba},
loosely bound molecular baryon-meson pentaquark states 
\cite{Karliner:2015ina,
Chen:2015loa,
Chen:2015moa,%
Roca:2015dva,%
He:2015cea,%
Huang:2015uda}, 
or peaks due to triangle-diagram 
processes~\cite{Guo:2015umn,
Meissner:2015mza,
Liu:2015fea,
Mikhasenko:2015vca}.

In this Letter, an analysis is presented of $\Lb\to\jpsi p\Km$ decays based on the combined data set collected by the LHCb collaboration in Run~1, with $pp$ collision energies of 7 and 8\tev corresponding to a total integrated luminosity of 3\invfb, 
and in Run~2 at 13\tev corresponding to 6\invfb. 
The \lhcb detector
is a single-arm forward spectrometer covering the pseudorapidity range \mbox{$2<\eta<5$}, 
described in detail in Refs.~\cite{LHCb-DP-2008-001,LHCb-DP-2014-002}. 
The data selection is similar to that used in Ref.~\cite{LHCb-PAPER-2015-029}.
However, in this updated analysis, the  
hadron-identification information is included in the
Boosted Decision Tree (BDT) discriminant,
which increases the $\Lb$ signal efficiency by almost a factor of two while leaving the background level  almost unchanged. 
The resulting sample contains $246$k $\Lb\to\jpsi p\Km$ decays (see the Supplemental Material to this Letter), which is nine times more than used in the Run~1 analyses~\cite{LHCb-PAPER-2015-029,LHCb-PAPER-2016-009}.

\begin{figure}[t]
    \begin{center}
        \includegraphics[width=0.49\linewidth]{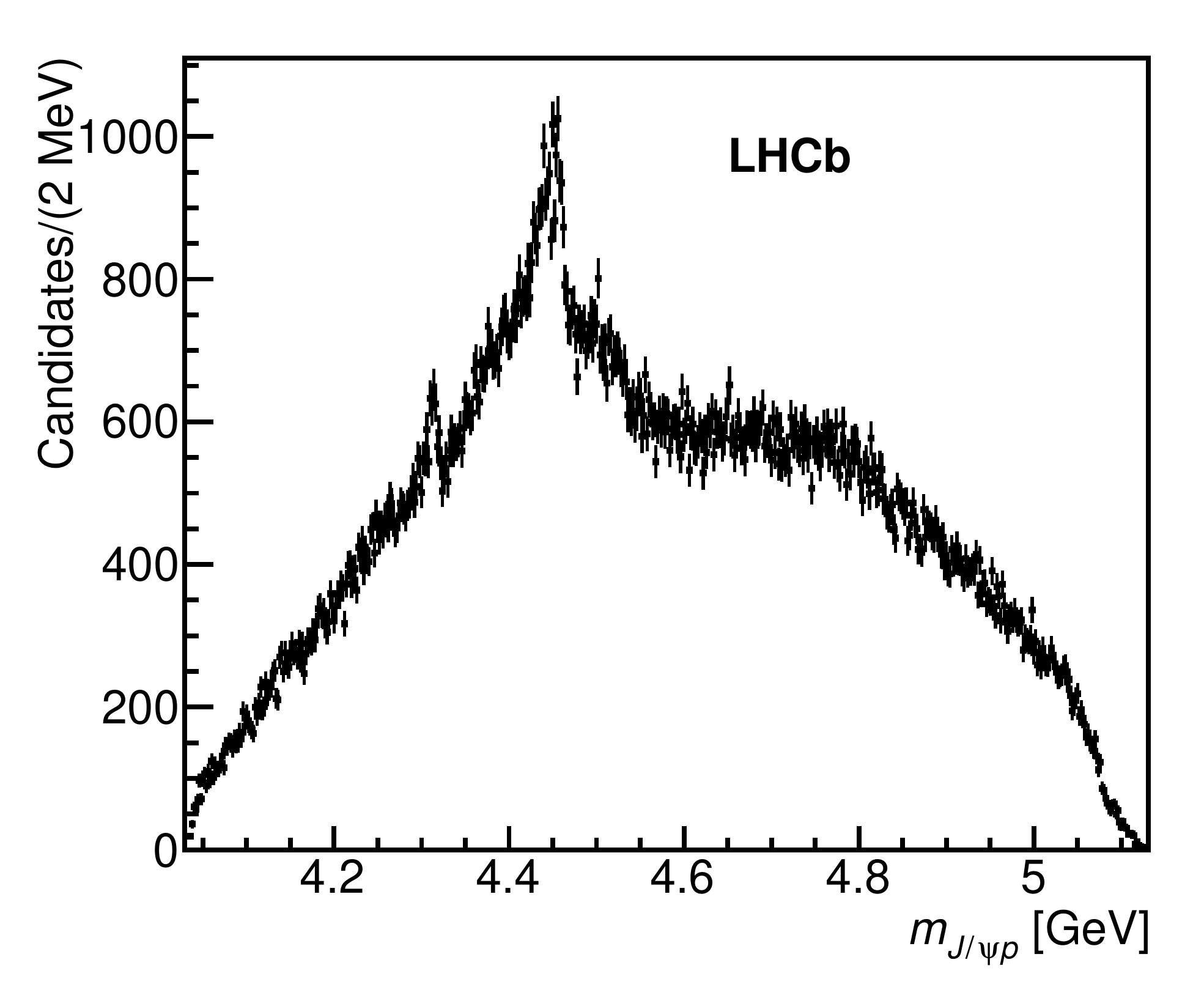}
    \includegraphics[width=0.49\linewidth]{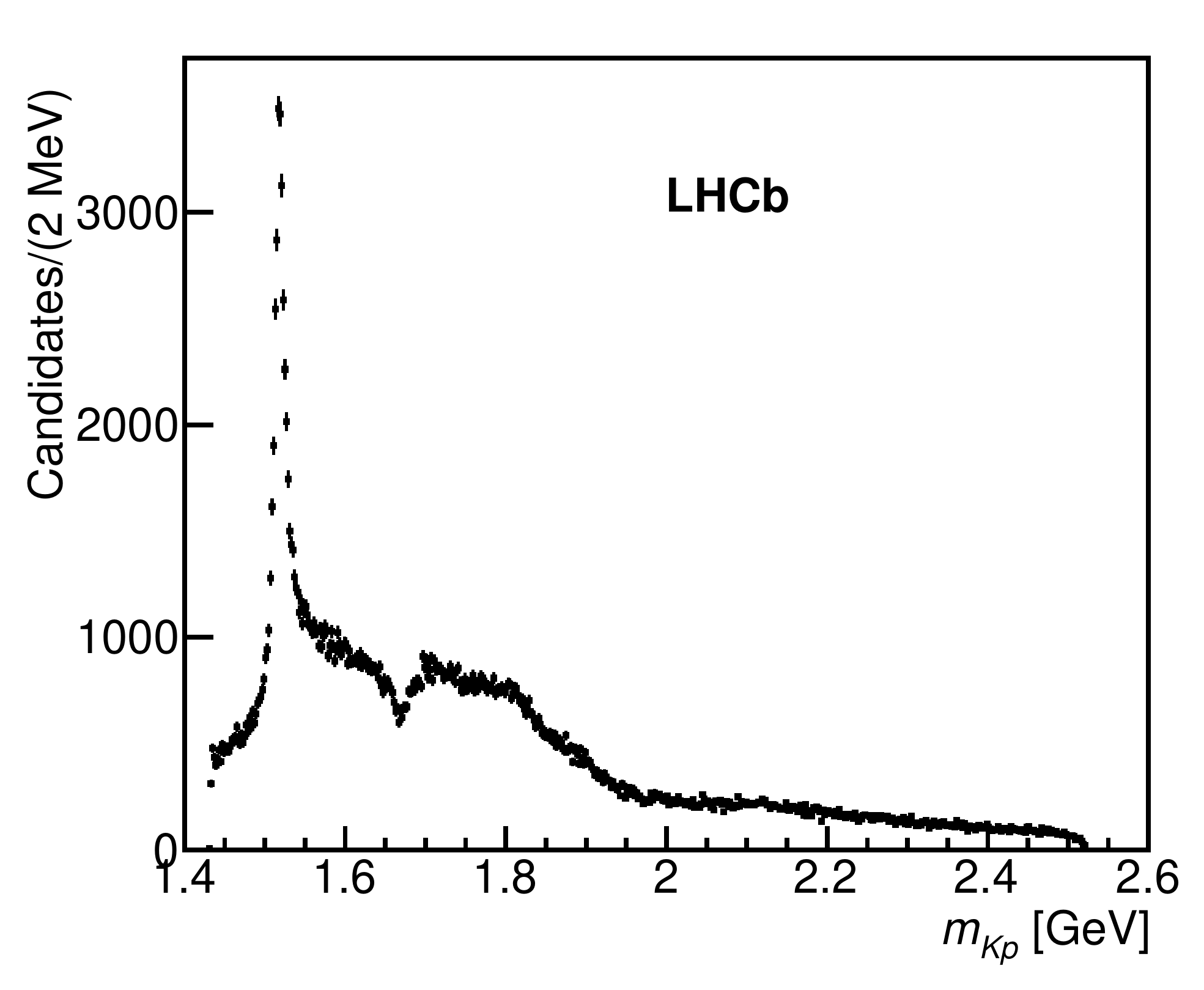}
    \end{center}
    \caption{Distribution of (left) $\mjpsip$ and (right) $\mkp$ for $\Lb\to\jpsi p\Km$ candidates. The prominent peak in $\mkp$ is due to the $\Lz(1520)$ resonance.}
    \label{fig:mjpsip_mkp}
\end{figure}

\begin{figure}[t]
    \centering
    \includegraphics[width=\linewidth]{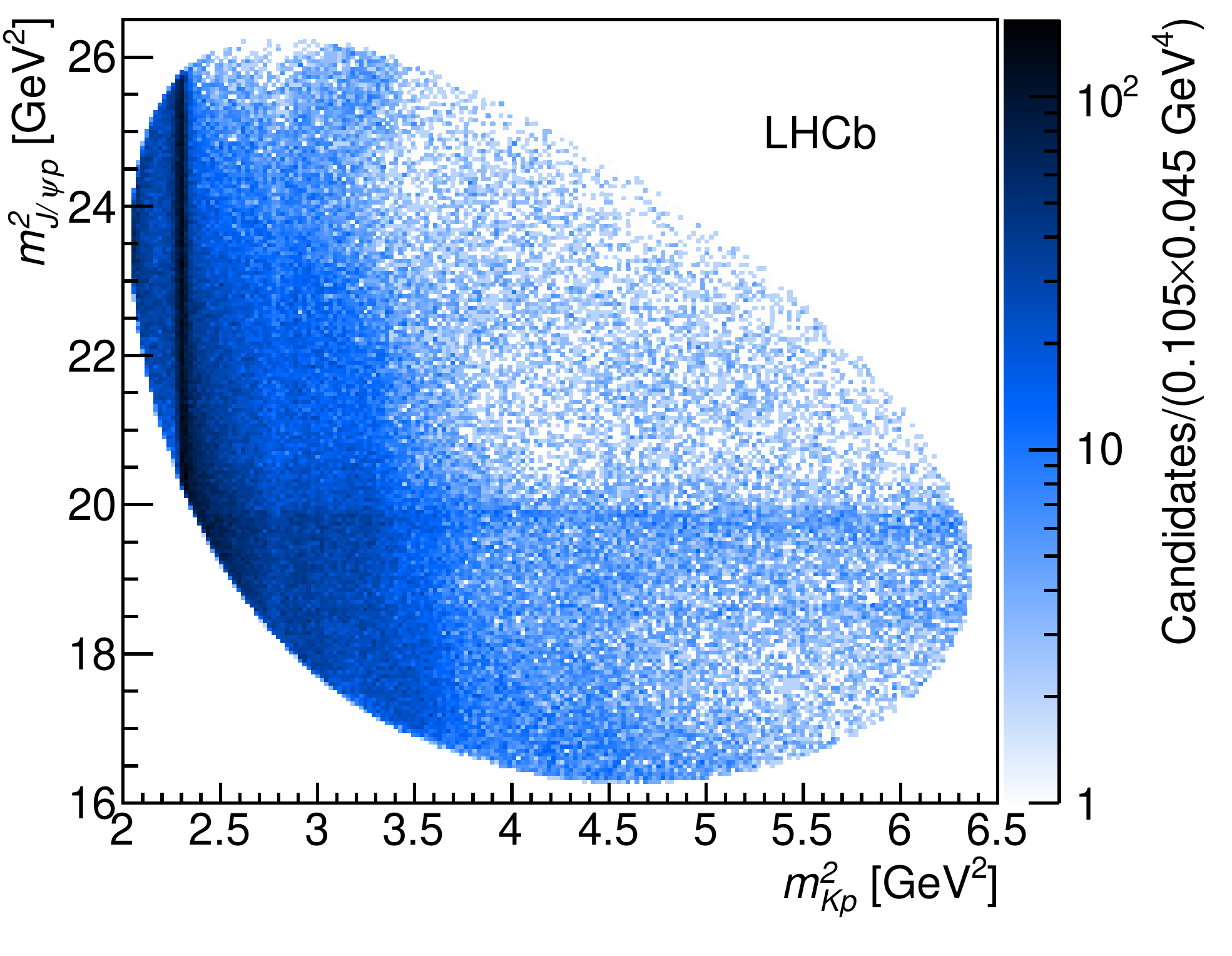}
    \caption{Dalitz plot of $\Lb\to\jpsi p\Km$ candidates. The data contain $6.4\%$ of non-$\Lb$ backgrounds, which are distributed smoothly over the phase-space. The vertical bands correspond to the $\Lz^*$ resonances. The horizontal bands correspond to the $P_c(4312)^+$, $P_c(4440)^+$, and $P_c(4457)^+$ structures at $m_{\jpsi p}^2 = 18.6$, $19.7$, and $19.9\gev^2$, respectively.}
    \label{fig:dalitz}
\end{figure}

\begin{figure}[t]
    \centering
    \includegraphics[width=\linewidth]{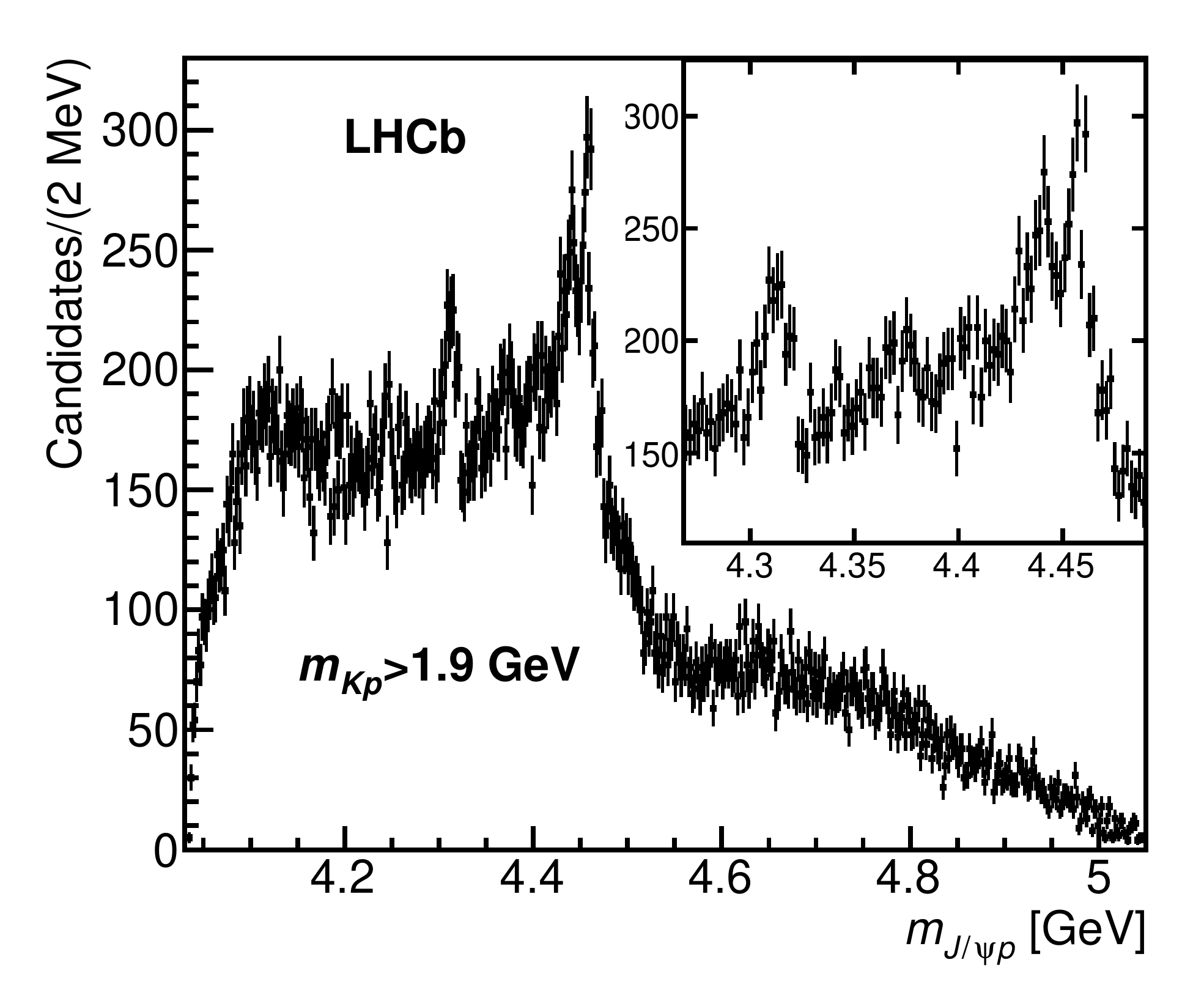}
    \caption{Distribution of $\mjpsip$ from $\Lb\to\jpsi p \Km$ candidates after  suppression of the dominant $\Lz^*\to p\Km$ contributions with the $\mkp>1.9$\gev requirement. The inset shows a zoom into the region of the narrow $P_c^+$ peaks.}
    \label{fig:mjpsi_spectrum_19}
\end{figure}

When this combined data set is fit with the same amplitude model used in Ref.~\cite{LHCb-PAPER-2015-029}, the $P_c(4450)^+$ and $P_c(4380)^+$ parameters are found to be consistent with the previous results. 
However, this should be considered only as a cross check, since analysis of this much larger data sample reveals additional peaking structures in the $\jpsi p$ mass spectrum, which are too small to have been significant before (see Fig.~\ref{fig:mjpsip_mkp} left).
A narrow peak is observed near 4312\mev with a width comparable to the mass resolution. The structure at 4450\mev is now resolved into two narrow peaks at 4440 and 4457\mev, which are more visible 
when the dominant $\Lz^*\to p \Km$ contributions, which peak at low $p \Km$ masses ($\mkp$) as shown in Fig.~\ref{fig:mjpsip_mkp}~right and Fig.~\ref{fig:dalitz}, are suppressed by requiring $\mkp>1.9$\gev (see Fig.~\ref{fig:mjpsi_spectrum_19}).  
This $\mkp$ requirement maximizes the expected signal significance for $P_c^+$ states that decay isotropically.

Performing a rigorous amplitude analysis of this new data sample is computationally challenging. The $\mjpsip$ mass resolution must be taken into account, and the size of the data sample to fit has greatly increased. 
Formulating an amplitude model whose systematic uncertainties are comparable to the statistical precision provided by this larger data sample is
difficult given the large number of $\Lz^*$ excitations  \cite{Capstick:1986bm,Loring:2001ky}, coupled-channel effects \cite{Fernandez-Ramirez:2015tfa},
and the possible presence of one or more wide $P_c^+$ contributions, like the previously reported $P_c(4380)^+$ state. Fortunately, the newly observed peaks are so narrow that it is not necessary to construct an amplitude model to prove that these states are not artifacts of interfering $\Lz^*$ resonances~\cite{LHCb-PAPER-2016-009}. 

Binned $\chi^2$ fits  are performed to the one-dimensional $\mjpsip$ distribution in the range $4.22 < \mjpsip < 4.57\gev$ to determine the masses ($M$), widths ($\Gamma$), and relative production rates (${\cal R}$) of the narrow $P_c^+$ states under the assumption that they  can be described by relativistic Breit--Wigner (BW) amplitudes.
These $\mjpsip$ fits alone cannot distinguish broad $P_c^+$ states from other contributions that vary slowly with $\mjpsip$. Therefore, a verification of the $P_c(4380)^+$ state observed in Ref.~\cite{LHCb-PAPER-2015-029}
awaits completion of an amplitude analysis of this new larger data set.

\begin{figure}[t]
    \centering
    \includegraphics[width=0.7\linewidth]{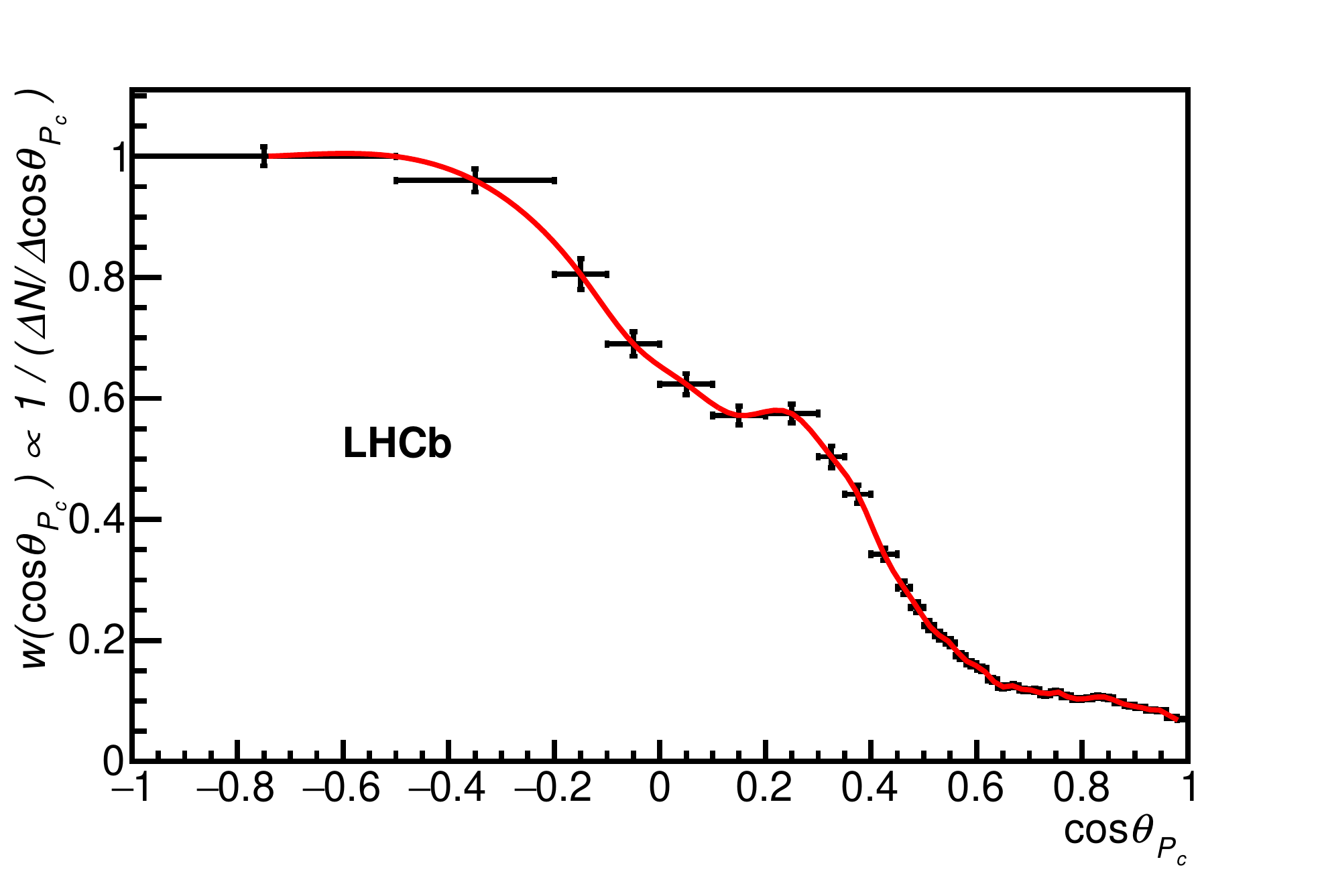}
    \vskip-0.5cm
    \caption{Weight function $w(\cos\theta_{Pc})$ applied to candidates, determined as the inverse of the density of $\Lb$ candidates in the narrow $P_c^+$ peak region. 
    The red line is a spline function used to interpolate between bin centers.}
    \label{fig:weight}
\end{figure}

Many variations of the $\mjpsip$ fits are performed to study the robustness of the measured $P_c^+$ properties.
The $\mjpsip$ distribution is fit both with and without requiring ${\mkp>1.9}$\gev, which removes over 80\% of the $\Lz^*$ contributions. 
In addition, fits are performed on the $\mjpsip$ distribution obtained by applying $\cos\theta_{P_c}$-dependent weights to each candidate to enhance the $P_c^+$ signal, where $\theta_{P_c}$ 
is the angle between the $\Km$ and $\jpsi$ in the $P_c^+$ rest frame (the $P_c^+$ helicity angle \cite{LHCb-PAPER-2015-029}).
The $\Lz^*$ contributions mostly populate the $\cos\theta_{P_c} > 0$ region. 
The weights are taken to be the inverse of the expected background at each $\cos\theta_{Pc}$, 
which is approximately given by the density of candidates observed in data since the signal contributions are small. 
The weight function 
is shown in Fig.~\ref{fig:weight}.
The best sensitivity to $P_c^+$ contributions is obtained from the $\cos\theta_{Pc}$-weighted $\mjpsip$ distribution, followed by the sample with the $\mkp>1.9$\gev requirement. However, since the background composition and shape are different in the three samples, the results from all three fits are used when assessing the systematic uncertainties.

The one-dimensional fit strategy is validated on ensembles of large simulated data sets sampled from several six-dimensional amplitude models, similar to those of Ref.~\cite{LHCb-PAPER-2015-029}, with or without a broad $P_c^+$ state and considering various $P_c^+$ quantum-number assignments. 
 The main conclusion from these studies is that
 the dominant systematic uncertainty is due to possible interference between various $P_c^+$ states. 
 Such interference effects cannot be unambiguously disentangled using the $\mjpsip$ distribution alone. 
 Therefore, fits are performed considering many possible interference configurations, with the observed variations in the $P_c^+$ properties assigned as systematic uncertainties.

In all fits, the $\mjpsip$ distribution is modeled by three narrow BW $P_c^+$ terms and a smooth parametrization of the background. 
Here, background refers to  $\Lz^*$ reflections, small non-$\Lb$ contributions (which comprise $6.4\%$ of the sample), and possibly additional broad $P_c^+$ structures.
 Many different background parametrizations are considered (discussed below), each of which is found to produce negligible bias in the $P_c^+$ parameters in the validation fits.
Each fit component is multiplied by a phase-space factor, $p\cdot q$, where $p$ ($q$) is the break-up momentum in the $\Lb\to P_c^+\Km$ ($P_c^+\to \jpsi p$) decay. 
Since the signal peaks are narrow, all fit components are convolved with the detector resolution, which is 2--3\mev in the fit region (see the Supplemental Material).  
Finally, the detection efficiency has negligible impact on the signal $\mjpsip$ distributions, and therefore, is not considered in these fits. 

\begin{figure}[t]
    \centering
    \includegraphics[width=0.9\linewidth]{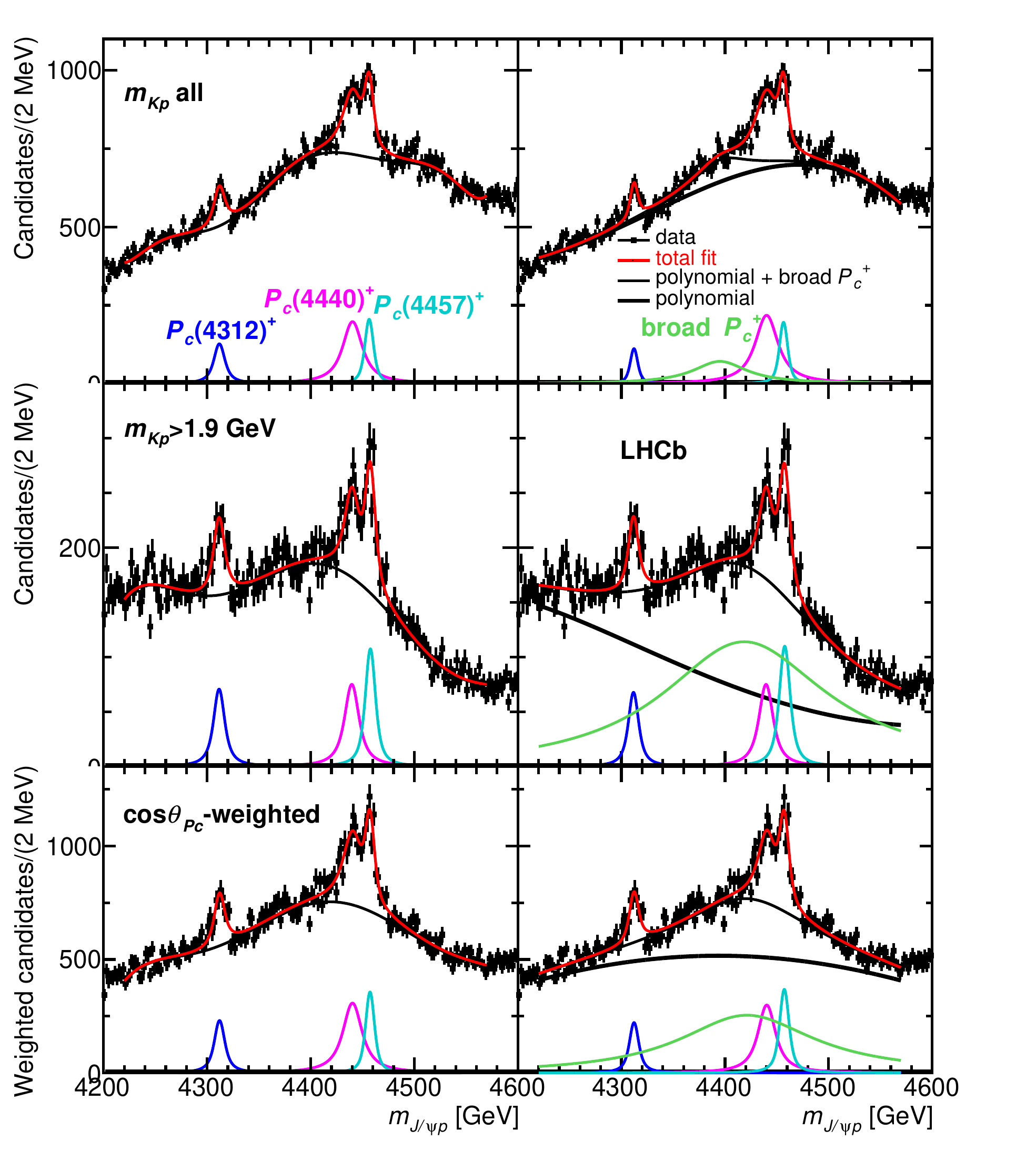}
        \caption{Fits to the $\mjpsip$ distributions of the (top row) inclusive, (middle row) $\mkp>1.9$\gev, and (bottom row) $\cos\theta_{Pc}$-weighted samples  with three incoherently summed BW amplitudes representing the narrow $P_c^+$ signals on top of a (left column) high-order polynomial function  or (right column) lower-order polynomial plus a broad $P_c^+$ state represented by a fourth BW amplitude. 
    } 
    \label{fig:incohfits}
\end{figure}

\begin{figure}[t]
    \centering
    \includegraphics[width=\linewidth]{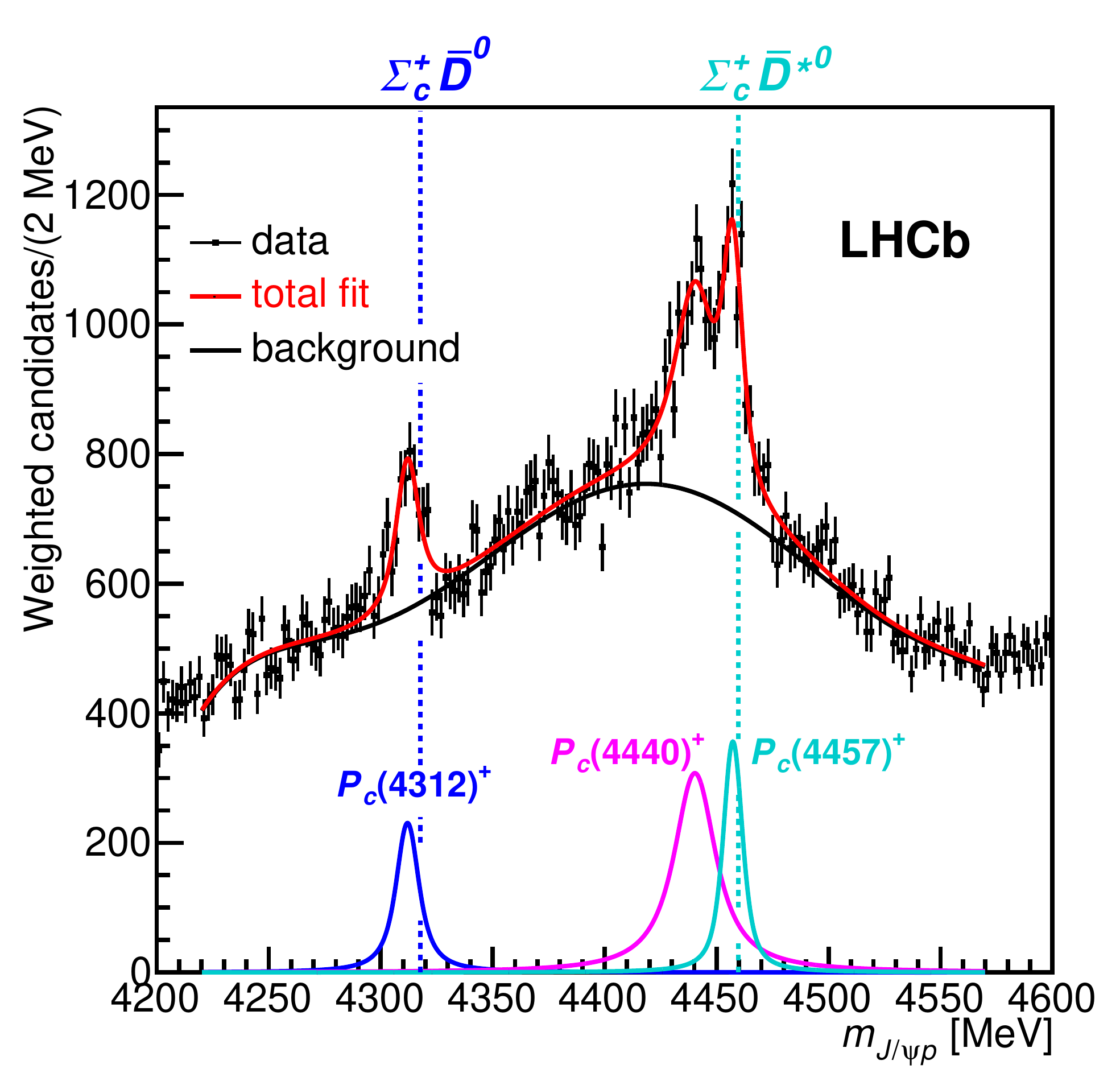}
    \caption{Fit to the $\cos\theta_{Pc}$-weighted $\mjpsip$ distribution with three BW amplitudes and a sixth-order polynomial background. This fit is used to determine the central values of the masses and widths of the $P_c^+$ states. The mass thresholds for the  $\Sigmac\Dzb$  and $\Sigmac\Dstarzb$ final states are superimposed.}
    \label{fig:defaultFit}
\end{figure}

In the nominal fits, the BW contributions are added incoherently. 
The results of these fits  are displayed in Fig.~\ref{fig:incohfits} for two parametrizations of the background: one using a high-order polynomial; and another using a low-order polynomial, along with an additional wide $P_c^+$ BW term whose mass and width are free to vary in the fits. 
For both background parametrizations, a range of polynomial orders is considered.
The lowest order used for each case is the smallest that adequately describes the data, which is found to correspond to the minimum order required to obtain unbiased $P_c^+$ estimators in the fit-validation studies in the absence of interference.
The highest orders are chosen such that the background model is capable of describing any structures that could be produced by either non-$P_c^+$ or broad-$P_c^+$ contributions.  
Figure~\ref{fig:defaultFit} shows the fit from which the central values of the $P_c^+$ properties are obtained, while the background-model-dependent variations observed in these properties are included in the systematic uncertainties. 
The fits with and without the broad $P_c^+$ state both describe the data well. Therefore, these fits can neither confirm nor contradict the existence of the $P_c(4380)^+$ state.

To determine the significance of the $P_c(4312)^+$ state, the change of the fit $\chi^2$ when adding this component is used as the test statistic, where the distribution under the null hypothesis is obtained from a large ensemble of pseudoexperiments.
The $p$-value, 
expressed in Gaussian standard deviations, corresponds to  $7.6\sigma$ ($8.5\sigma$) for the fits to the $\mkp>1.9$\gev ($\cos\theta_{Pc}$-weighted) distribution, ignoring the look-elsewhere effect. 
To account for this effect, 
the $\mjpsip$ distribution in each pseudoexperiment is scanned to find the most significant narrow and isolated peak (excluding the 4450\mev peak region). 
This method lowers the  $P_c(4312)^+$ significance to $7.3\sigma$ ($8.2\sigma$).

To evaluate the significance of the two-peak structure versus the one-peak interpretation of the 4450\mev region,
the null hypothesis uses just one BW to encompass both the $P_c(4440)^+$ and $P_c(4457)^+$ peaks (the fit also includes the $P_c(4312)^+$ BW), which gives $P_c(4450)^+$ mass and width values that are consistent with those obtained from the amplitude analysis of Ref.~\cite{LHCb-PAPER-2015-029}. 
Pseudoexperiments are again used to determine the $\Delta\chi^2$ distribution under the null hypothesis. The significance of the two-peak structure is 
$5.4\sigma$ ($6.2\sigma$) for the $\mkp>1.9$\gev ($\cos\theta_{Pc}$-weighted) samples. 
This significance is large enough to render the single peak interpretation of the 4450\mev region obsolete. 
Therefore, the results presented here for this structure supersede those previously presented in Ref.~\cite{LHCb-PAPER-2015-029} (see the Supplemental Material for more detailed discussion). 
To investigate the systematic uncertainties on $P_c^+$ properties due to interference, which can only be important for $P_c^+$ resonances with the same spin and parity, fits to the $\cos\theta_{Pc}$-weighted distribution are repeated using various coherent sums of two of the BW amplitudes. 
Each of these fits includes a phase between interfering resonances as an extra free parameter. 
None of the interference effects studied is found to produce a significant $\Delta\chi^2$ relative to the fits using an incoherent sum of BW amplitudes. However, substantial shifts in the $P_c^+$ properties are observed, and are included in the systematic uncertainties. 
For example, in such fit the $P_c(4312)^+$ mass increases, while its width is rather stable, leading to 
a large positive systematic uncertainty of $6.8$\mev on its mass.

As in Ref.~\cite{LHCb-PAPER-2015-029}, the $\Lb$ candidates are kinematically constrained to the known $\jpsi$ and $\Lb$ masses~\cite{Tanabashi:2018oca}, which substantially improves the $\mjpsip$ resolution and determines the absolute mass scale with an accuracy of $0.2$\mev.
The mass resolution is known with a 10\% relative uncertainty.
Varying this within its uncertainty changes the widths of the narrow states in the nominal fit by up to 0.5\mev, 0.2\mev, and 0.8\mev for the $P_c(4312)^+$, $P_c(4440)^+$, and $P_c(4457)^+$ states, respectively.
The widths of all three narrow $P_c^+$ peaks are consistent with the mass resolution within the systematic uncertainties. 
Therefore, upper limits are placed on their natural widths at the 95\% confidence level (CL), which account for the uncertainty on the detector resolution and in the fit model. 

A number of additional fits are performed when evaluating the systematic uncertainties.
The nominal fits assume S-wave (no angular momentum) production and decay. Including P-wave factors in the BW amplitudes has negligible effect on the results.
In addition to the nominal fits with three narrow peaks in the $4.22 < \mjpsip < 4.57\gev$ region, 
fits including only the $P_c(4312)^+$ are performed in the narrow 4.22--4.44\gev range.
Fits are also performed using a data sample selected with an alternative approach, where no BDT is used resulting in about twice as much background. 

The total systematic uncertainties assigned on the mass and width of each narrow $P_c^+$ state are taken to be the largest deviations observed among all fits.
These include the fits to all three versions of the $\mjpsip$ distribution, 
each configuration of the $P_c^+$ interference,
all variations of the background model,
and each of the additional fits just described.
The masses, widths, and the relative contributions (${\cal R}$ values) of the three narrow $P_c^+$ states, 
including all systematic uncertainties, are given in Table~\ref{tab:results}.

\long\def\resultstable{
\renewcommand*{\arraystretch}{1.3}
\def\1{\phantom{1}}
    \centering
       \caption{Summary of $P_c^+$ properties. The central values are based on the fit displayed in Fig.~\ref{fig:defaultFit}.}
    \begin{tabular}{c|c|cc|c}
    State &  $M$ [\mev] & 
    $\Gamma$ [\mev] & (95\% CL) & ${\cal R}$ [\%] \\
    \hline
  $P_c(4312)^+$ &  $4311.9\pm0.7^{+6.8}_{-0.6}$ &
          \phantom{2}$9.8\pm2.7^{+\1 3.7}_{-\1 4.5}$ & $(<27)$  &  $0.30\pm0.07^{+0.34}_{-0.09}$  \\
  $P_c(4440)^+$ &  $4440.3\pm1.3^{+4.1}_{-4.7}$ &
          $20.6\pm4.9_{-10.1}^{+\1 8.7}$ & $(<49)$ & $1.11\pm0.33^{+0.22}_{-0.10}$\\
  $P_c(4457)^+$ &  $4457.3\pm0.6^{+4.1}_{-1.7}$ &
          \phantom{2}$6.4\pm2.0_{-\1 1.9}^{+\1 5.7}$ & $(<20)$ & $0.53\pm0.16^{+0.15}_{-0.13}$ \\
    \end{tabular}
    \label{tab:results}
}

\ifthenelse{\boolean{prl}}{
\begin{table*}[t]
\resultstable
\end{table*}
}{
\begin{table}[t]
\resultstable
\end{table}
}

To obtain estimates of the relative contributions of the $P_c^+$ states, the $\Lb$ candidates are weighted by the inverse of the reconstruction efficiency, which is parametrized in all six dimensions of the $\Lb$ decay phase-space 
(Eq.~(68) in the \href{https://journals.aps.org/prl/supplemental/10.1103/PhysRevLett.115.072001/PRL-Pentaquark-suppl.pdf}{Supplemental Material} to Ref.~\cite{LHCb-PAPER-2016-029}).
The efficiency-weighted $\mjpsip$ distribution, without the $\mkp > 1.9\gev$ requirement, is fit to determine the $P_c^+$ contributions, which 
are then divided by the efficiency-corrected and background-subtracted $\Lb$ yields.
This method makes the results independent of the unknown quantum numbers and helicity structure of the $P_c^+$ production and decay. 
Unfortunately, this approach also suffers from  large  $\Lz^*$ backgrounds and from sizable fluctuations in the low-efficiency regions. 
In these fits, the $P_c^+$ terms are added incoherently, absorbing any interference effects, which can be large (see, {\em e.g.}, Fig.~S2 
in the Supplemental Material), into the BW amplitudes.
Therefore, the \mbox{${\cal R}\equiv {{\cal B}(\Lb\to P_c^+\Km){\cal B}(P_c^+\to\jpsi p)}/{{\cal B}(\Lb\to\jpsi p\Km)}$} values reported for each $P_c^+$ state differ from the fit fractions typically reported in amplitude analyses, since  $\mathcal{R}$ includes both the BW amplitude squared and all of its interference terms.  
Similar fit variations are considered here as above, {\em e.g.}, different background models and selection criteria are all evaluated. 
The resulting systematic uncertainties on 
${\cal R}$
are large, 
as shown in Table~\ref{tab:results}.

The narrow widths of the $P_c^+$ peaks make a compelling case for the bound-state character of the observed states. 
However, it has been pointed out by many authors 
\cite{Guo:2015umn,
Meissner:2015mza,
Liu:2015fea,
Mikhasenko:2015vca}
that peaking structures in this $\jpsi p$ mass range can also be generated by triangle diagrams. The $P_c(4312)^+$ and $P_c(4440)^+$ peaks are unlikely to arise from triangle diagrams, due to a lack of any appropriate hadron-rescattering thresholds as discussed in more detail in the Supplemental Material. 
The $P_c(4457)^+$ peaks at the $\Lc(2595)\Dzb$ threshold ($J^P=1/2^+$ in S-wave)~\cite{Liu:2015fea}, and  
the $D_{s1}(2860)^-$ meson is a suitable candidate to be exchanged in the corresponding triangle diagram. 
However, this triangle-diagram term does not describe the data nearly as well as the BW does (Fig.~S5 
in the Supplemental Material \cite{SupplPRL}). 
This possibility deserves more scrutiny within the amplitude-analysis approach. 

Narrow $P_c^+$ states could arise by binding a narrow baryon with a narrow meson, where the separation of $c$ and $\bar{c}$ into distinct confinement volumes provides a natural suppression mechanism for the $P_c^+$ widths.  
The only narrow baryon-meson combinations with mass thresholds in the appropriate mass range are $p\chi_{cJ}$, $\Lc\bar{D}^{(*)0}$, and $\Sigmares_{c}\bar{D}^{(*)}$ 
(both $\Sigmac\bar{D}^{(*)0}$ and $\Sigmares_c^{++}\bar{D}^{(*)-}$ are possible, the threshold for the latter is about 5\mev higher than the former). 
There is no known S-wave binding mechanism for $p\chi_{cJ}$ combinations~\cite{Burns:2015dwa}
and $\Lc\bar{D}^{(*)0}$ interactions are expected to be repulsive, leaving only the $\Sigmares_{c}\bar{D}^{(*)}$ pairs expected to form bound states~\cite{Wang:2011rga,Yang:2011wz,Wu:2012md}. 
The masses of the $P_c(4312)^+$ and $P_c(4457)^+$ states are approximately $5\mev$ and $2\mev$ below the $\Sigmac\Dzb$ and $\Sigmac\Dstarzb$ thresholds, respectively, as illustrated in Fig.~\ref{fig:defaultFit}, making them excellent candidates for bound states of these systems. 
The $P_c(4440)^+$ could be the second $\Sigmares_{c}\Dstarb$ state, with about $20\mev$ of binding energy, since two states with $J^P=1/2^-$ and $3/2^-$ are possible.
In fact, several papers on hidden-charm states created dynamically by charmed meson-baryon interactions \cite{Wu:2010jy,Wu:2010vk,Xiao:2013yca} were published well before the first observation of the $P_c^+$ structures~\cite{LHCb-PAPER-2015-029}~and some of these predictions for $\Sigmac\Dzb$ and $\Sigmac\Dstarzb$ states~\cite{Wang:2011rga,Yang:2011wz,Wu:2012md} are consistent with the observed narrow $P_c^+$ states. 
Such an interpretation of the $P_c(4312)^+$ state (implies $J^P=1/2^-$) would point to the importance of $\rho$-meson exchange,
since a pion cannot be exchanged in this system \cite{Karliner:2015ina}.

In summary, the nine-fold increase in the number of $\Lb\to\jpsi p\Km$ decays reconstructed with the \lhcb detector sheds more light onto the $\jpsi p$ structures found in this final state.
The previously reported $P_c(4450)^+$ peak \cite{LHCb-PAPER-2015-029} is confirmed and resolved at 5.4$\sigma$ significance into two narrow states: the $P_c(4440)^+$ and $P_c(4457)^+$ exotic baryons.
A narrow companion state, $P_c(4312)^+$, is discovered with 7.3$\sigma$ significance.

The minimal quark content of these states is $duuc\bar{c}$.
Since all three states are narrow and below the $\Sigmac\Dzb$ and $\Sigmac\Dstarzb$ ($[duc][u\bar{c}]$) thresholds within plausible hadron-hadron binding energies,
they provide the strongest experimental evidence to date for the existence of bound states of a baryon and a meson.
The $\Sigmac\Dzb$ ($\Sigmac\Dstarzb$) threshold is within the extent of the $P_c(4312)^+$ ($P_c(4457)^+$) peak, 
and therefore virtual \cite{PhysRev.134.B1307}
rather than bound states are among the plausible explanations.  
In simple tightly bound pentaquark models, the proximity of these states to baryon-meson thresholds would be coincidental, and furthermore, it is difficult to accommodate their narrow widths~\cite{Hiyama:2018ukv}.
A potential barrier between diquarks, which could separate the $c$ and $\bar{c}$ quarks, has been proposed to solve similar difficulties for tetraquark candidates~\cite{Maiani:2017kyi}.
An interplay between tightly bound pentaquarks and the $\Sigmares_c\Db$, $\Sigmares_c\Dstarb$ thresholds may also be responsible for the $P_c^+$ peaks
\cite{Bugg:2008wu,Guo:2014iya,Blitz:2015nra,Guo:2016bjq}. 
Therefore, such 
alternative explanations cannot be ruled out. 
Proper identification of the internal structure of the observed states will require more experimental and theoretical scrutiny.

\section*{Acknowledgements}
%
%
\noindent
We express our gratitude to our colleagues in the CERN accelerator departments for the excellent performance of the LHC. We thank the technical and administrative staff at the LHCb institutes.
We acknowledge support from CERN and from the national agencies:
CAPES, CNPq, FAPERJ and FINEP (Brazil);
MOST and NSFC (China);
CNRS/IN2P3 (France);
BMBF, DFG and MPG (Germany);
INFN (Italy);
NWO (Netherlands);
MNiSW and NCN (Poland);
MEN/IFA (Romania);
MSHE (Russia);
MinECo (Spain);
SNSF and SER (Switzerland);
NASU (Ukraine);
STFC (United Kingdom);
NSF (USA).
We acknowledge the computing resources that are provided by CERN, IN2P3 (France), KIT and DESY (Germany), INFN (Italy), SURF (Netherlands), PIC (Spain), GridPP (United Kingdom), RRCKI and Yandex LLC (Russia), CSCS (Switzerland), IFIN-HH (Romania), CBPF (Brazil), PL-GRID (Poland) and OSC (USA).
We are indebted to the communities behind the multiple open-source software packages on which we depend.
Individual groups or members have received support from AvH Foundation (Germany); EPLANET, Marie Sk\l{}odowska-Curie Actions and ERC (European Union); ANR, Labex P2IO and OCEVU, and R\'{e}gion Auvergne-Rh\^{o}ne-Alpes (France); Key Research Program of Frontier Sciences of CAS, CAS PIFI, and the Thousand Talents Program (China); RFBR, RSF and Yandex LLC (Russia); GVA, XuntaGal and GENCAT (Spain); the Royal Society and the Leverhulme Trust (United Kingdom); Laboratory Directed Research and Development program of LANL (USA); CONACYT (Mexico).

\newpage 
 


\clearpage
\newpage

\begin{center}
  \textbf{\large\normalfont\bfseries\boldmath\papertitle } \\
\vspace{0.05in}
{ \it \large Supplemental Material}\\
\vspace{0.05in}
\end{center}

\setcounter{equation}{0}
\setcounter{figure}{0}
\setcounter{table}{0}
\setcounter{section}{0}
\renewcommand{\theequation}{S\arabic{equation}}
\renewcommand{\thefigure}{S\arabic{figure}}
\renewcommand{\thetable}{S\arabic{table}}
\newcommand\ptwiddle[1]{\mathord{\mathop{#1}\limits^{\scriptscriptstyle(\sim)}}}

\section{\boldmath $\Lb\to\jpsi p\Km$ candidates}
\label{sec:Lbsample}

\begin{figure}[b]
  \begin{center}
     \includegraphics[width=\linewidth]{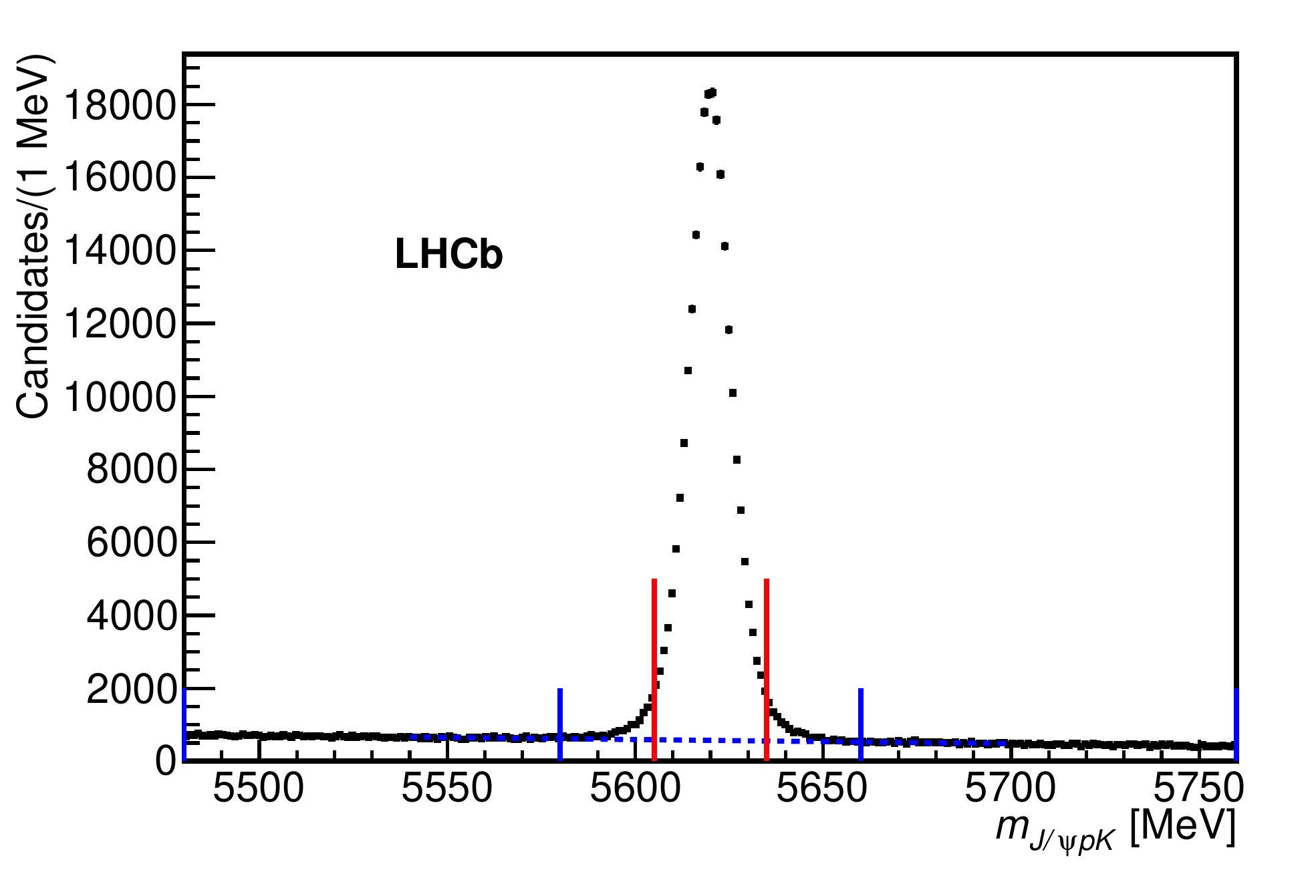}
  \end{center}
  \caption{
  Invariant mass spectrum of $\jpsi p\Km$ candidates.
  The $\Lb$ signal region  is between the vertical red lines.
  A linear interpolation of the background, determined from the sideband regions (bounded by the shorter vertical blue lines), to the signal region is shown by the dashed blue line. 
  }
  \label{fig:Lbsample}
\end{figure}

The $\Lb\to\jpsi p\Km$ sample analyzed in the Letter is selected by requiring that the invariant mass of $\jpsi p\Km$ candidates is in the 5605--5635\mev range. To determine the $\Lb$ signal yield within this range, the background density is linearly interpolated from the  5480--5580\mev and 5660--5760\mev sidebands to the signal region, as illustrated in Fig.~\ref{fig:Lbsample}.
There are $246$k $\Lb$ decays with 6.4\%\ background contamination in the analyzed sample.

After selecting candidates in the $\Lb$ signal region indicated in Fig.~\ref{fig:Lbsample}, the $\Lb$ mass constraint is imposed on all $\Lb$ candidates, in addition to the $\jpsi$ mass and vertex constraints, to improve the $\mjpsip$  resolution.    
To a good approximation, the mass resolution is Gaussian with a  standard deviation ($\sigma_m$) given by
\begin{equation}
\sigma_m(\mjpsip) = \left[ 2.71 - 6.56\cdot10^{-6} (\mjpsip/\mev - 4567)^2\right]\mev.
\end{equation}


\section{Example fit with interference}\label{sec:interference}
\begin{figure}[h]
    \centering
    \includegraphics[width=\linewidth]{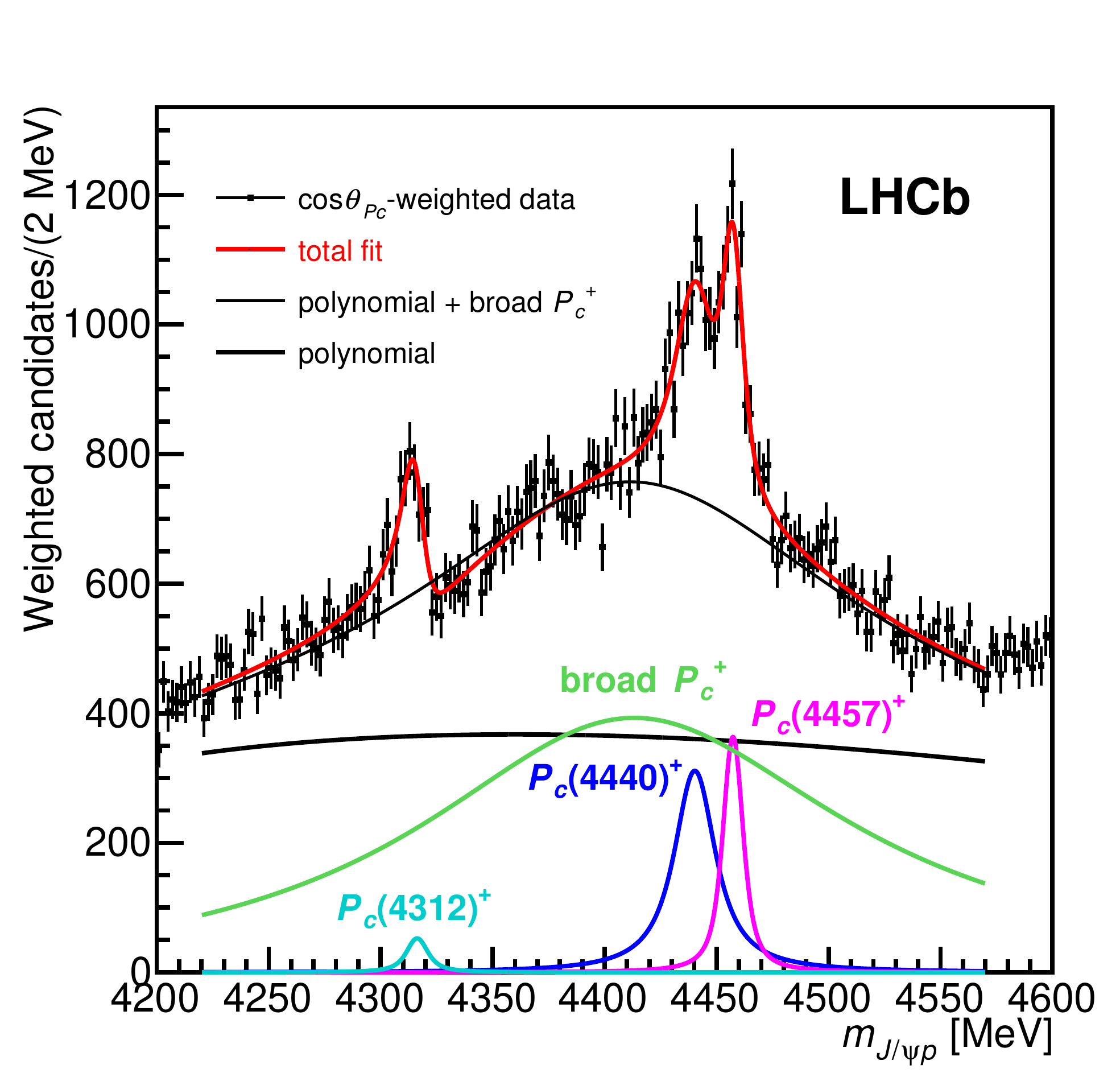}
    \caption{Fit to the $\cos\theta_{Pc}$-weighted $\mjpsip$ distribution with four BW amplitudes and a linear background. The broad $P_c^+$ state is added coherently to the $P_c(4312)^+$ amplitude. 
    In this fit model, the magnitude of the $P_c(4312)^+$ peak in the data is dominated by its interference with the broad $P_c^+$ state.
    Each $P_c^+$ contribution is displayed as the BW amplitude squared (the interference contributions are not shown individually). 
    }
    \label{fig:cohfit}
\end{figure}

Figure~\ref{fig:cohfit} shows an example fit with interfering resonances.

\quad

\section{Study of triangle diagrams}
\label{sec:triangles}

\begin{figure}[t]
  \begin{center}
     \includegraphics[width=\linewidth]{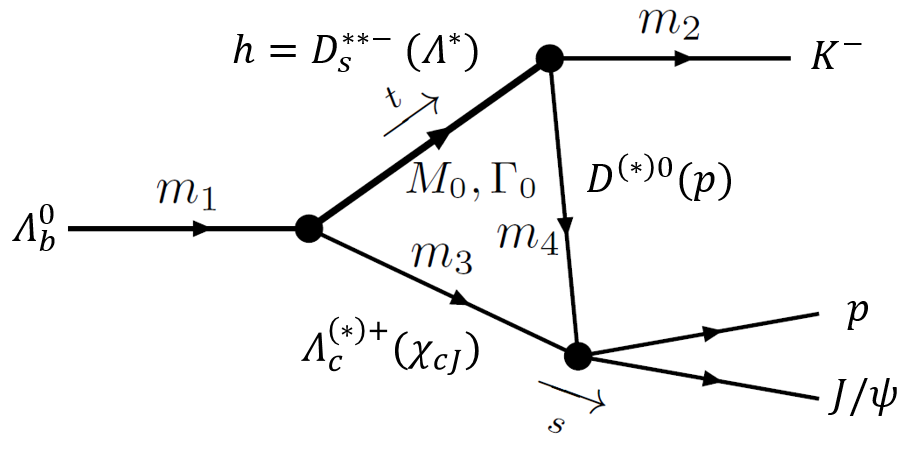}
    \vspace*{-0.7cm}
  \end{center}
  \caption{
  Triangle diagram for the $\Lb\to\jpsi p\Km$ decay. The figure defines the symbols used in the formulae in the text.
  }
  \label{fig:triangle}
\end{figure}

The narrow widths of the $P_c^+$ peaks make a compelling case for the bound-state character of the observed states. 
However, it has been pointed out by many authors 
\cite{Guo:2015umn,
Meissner:2015mza,
Liu:2015fea,
Mikhasenko:2015vca}
that peaking structures in this $\jpsi p$ mass range can also be generated by triangle diagrams (see Fig.~\ref{fig:triangle}).
In these processes, the $\Lb$ baryon (of mass $m_1$) decays into two nearly on-mass-shell hadrons, one of which (of mass $m_h \equiv \sqrt{t}$) is an excited strange meson or baryon (denoted  here as $h$) that subsequently emits a kaon (of mass $m_2$) and a non-strange decay product (of mass $m_4$) that rescatters with the other $\Lb$ child (of mass $m_3$) into the $\jpsi p$ system (of $\mjpsip \equiv \sqrt{s}$). Such triangle-diagram processes are known to peak when all three hadrons in the triangle are nearly on their mass shells.
Since the overall probability across coupled channels must be conserved, a peak in the $\jpsi p$ channel is only possible if there is a corresponding depletion in the final state composed of the particles that rescatter in Fig.~\ref{fig:triangle} to form the $\jpsi p$~\cite{Szczepaniak:2015hya}.

The triangle-diagram contribution often peaks at a threshold, given by the sum of the masses of the rescattering hadrons ($m_3+m_4$) creating a cusp. For a fine-tuned BW resonance mass of the intermediate hadron $h$ ($M_0$), the rate can peak above (but never below) the corresponding threshold.
The amplitude for a triangle-diagram process, which incorporates a finite width for the exchanged particle ($\Gamma_0$), is given by 
\begin{equation}
\label{eq:triangle}
    A(s) \propto \int_{(m_2+m_4)^2}^{\infty} dt\, \left| {\rm BW}(t|M_0,\Gamma_0) \right|^2\, F(t,s), 
\end{equation}
where all quantities are defined in Fig.~\ref{fig:triangle}. 
The BW term corresponds to the exchanged $h$ hadron.
The Feynman triangle-amplitude formula is expressed in terms of a one-dimensional integral over a single Feynman parameter $x$ as follows:
\begin{equation}
    F(t,s)\equiv2\int_0^1 \frac{dx}{y_- - y_+}\left[ \log\left(1-\frac{2sx}{y_+}\right) - \log\left(1-\frac{2sx}{y_-}\right)\right],
\end{equation}
where
\ifthenelse{\boolean{prl}}{%
\begin{align}
    y_{\pm}\equiv & y_{\pm}(s,t,x) \equiv  (m_1^2-m_2^2+s)x - m_1^2 + m_2^2 + m_3^2 - m_4^2 \notag\\
    & \pm\left\{i\epsilon + x^2\lambda(s,m_1^2,m_2^2)  \right. \notag\\
    & + 2x\left[(m_1^2+m_2^2-m_3^2-m_4^2+2t)s  \right. \notag\\ 
    & \left.  - (m_1^2-m_2^2) (m_1^2-m_2^2-m_3^2+m_4^2)\right] \notag\\
    & \left. -4st + (m_1^2-m_2^2-m_3^2+m_4^2)^2\right\}^{1/2}.
\end{align}
}{%
\begin{align}
    y_{\pm}\equiv & y_{\pm}(s,t,x) \equiv  (m_1^2-m_2^2+s)x - m_1^2 + m_2^2 + m_3^2 - m_4^2 \notag\\
    & \pm\left\{i\epsilon + x^2\lambda(s,m_1^2,m_2^2) + 2x\left[(m_1^2+m_2^2-m_3^2-m_4^2+2t)s  \right. \right. \notag\\ 
    & \left. \left. - (m_1^2-m_2^2) (m_1^2-m_2^2-m_3^2+m_4^2)\right]  -4st + (m_1^2-m_2^2-m_3^2+m_4^2)^2\right\}^{1/2}.
\end{align}
}
Here, $\lambda(a,b,c)=a^2+b^2+c^2-2ab-2ac-2bc$ and $\epsilon$ is a small real number.

\begin{figure}[htbp]
    \centering
    \includegraphics[width=\linewidth]{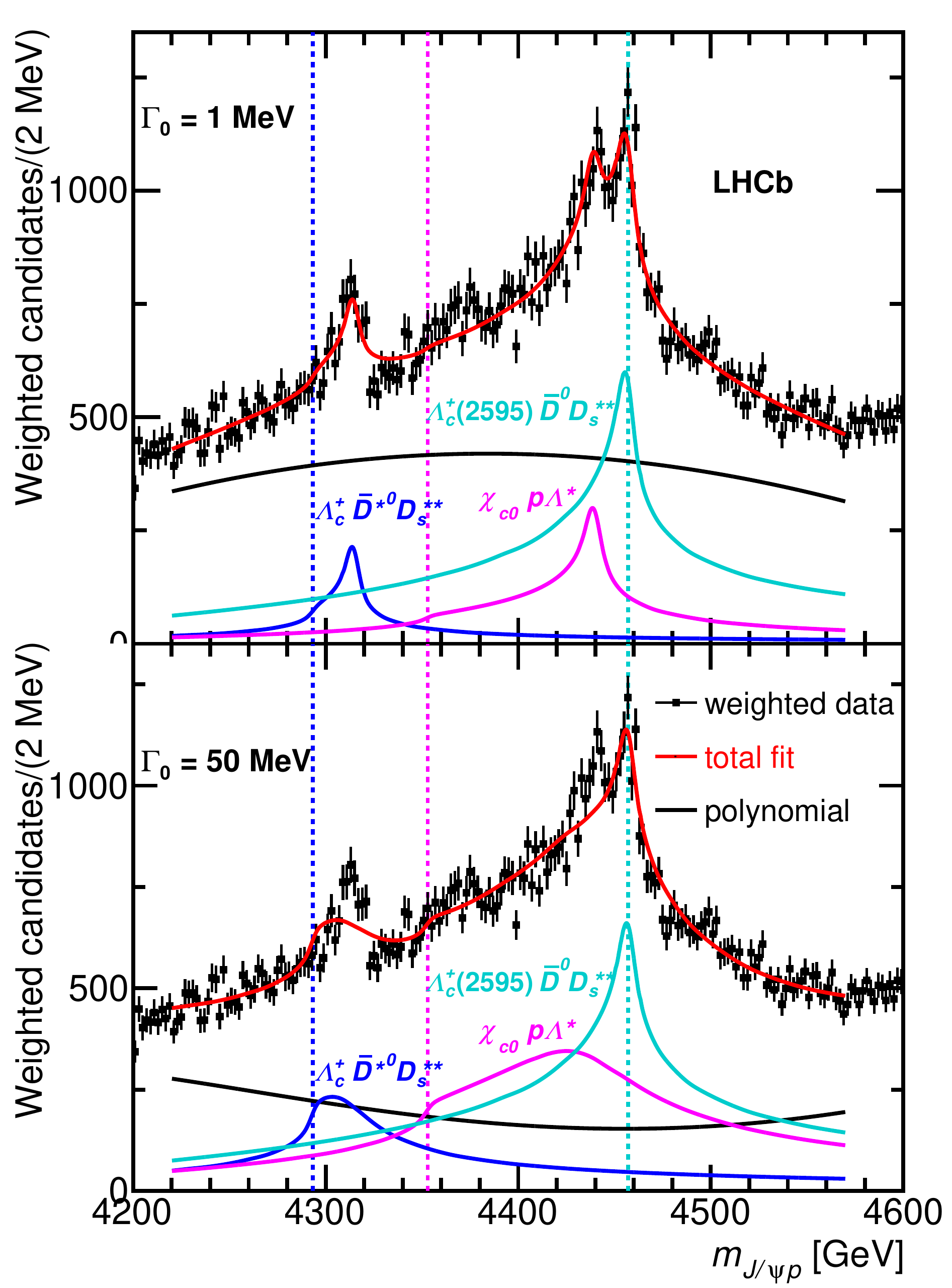}
    \caption{Fit 
    of three triangle-diagram amplitudes and a quadratic background to the $\cos\theta_{Pc}$-weighted
    distribution.
    The widths of the excited particles exchanged in the triangles is (top)~an unrealistic value of 1\mev or (bottom) a more plausible value of 50\mev.  
    Individual triangle diagram contributions are also shown.
    The dashed vertical lines are the $\Lc\Dstarzb$, 
    $\chi_{c0}p$ and $\Lc(2595)\Dzb$  thresholds.
    }
    \label{fig:triangleFits}
\end{figure}

\begin{figure}[htbp]
    \centering
    \includegraphics[width=\linewidth]{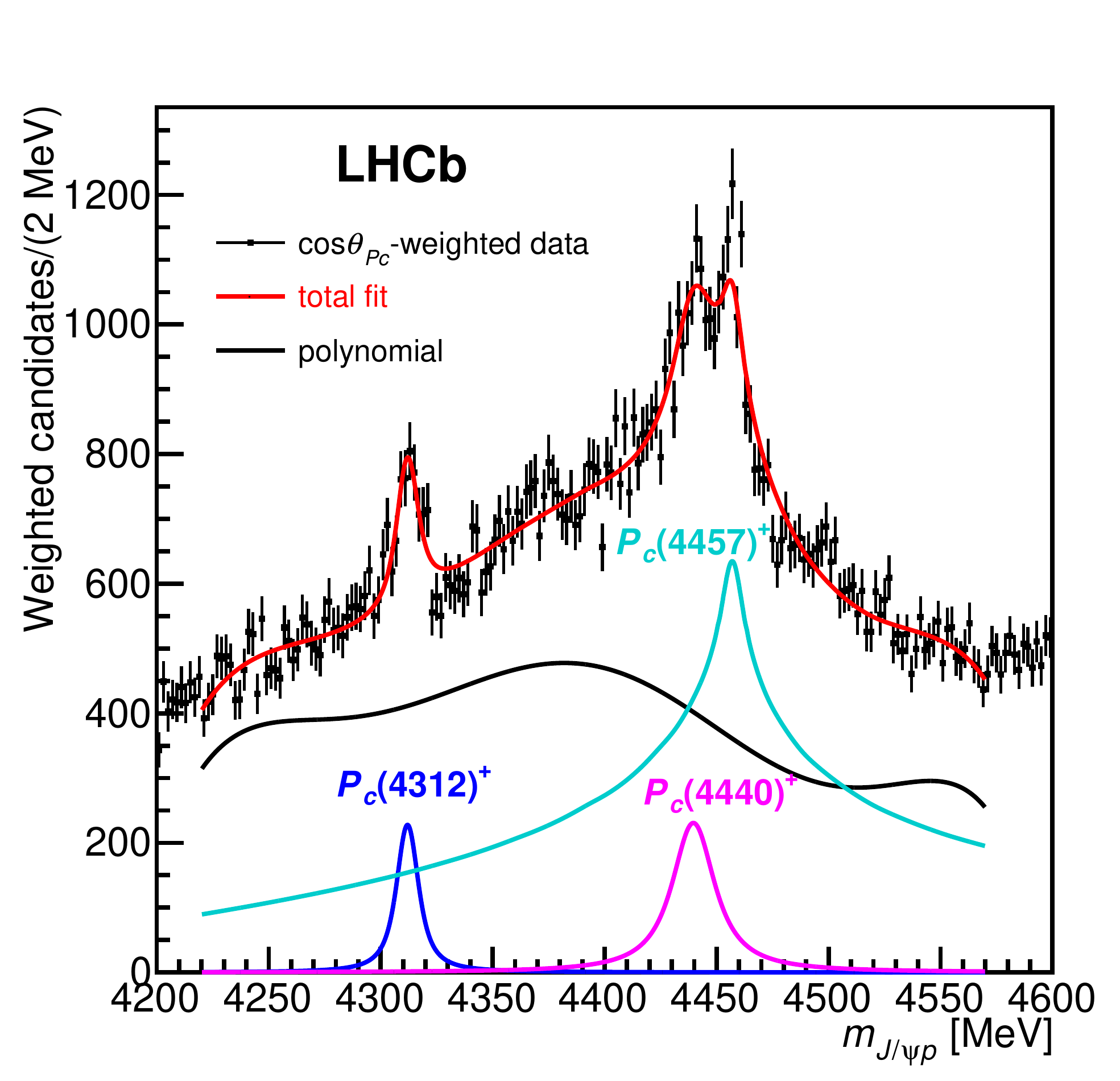}
    \caption{Fit 
    of two BW amplitudes, one triangle-diagram amplitude, 
    and a sixth-order polynomial background to the $\cos\theta_{Pc}$-weighted
    distribution.
    The width of the excited $D_s^*$ state exchanged in the triangle loop is 
    set to $\Gamma(D_{s1}(2860)^-)=159$\mev~\cite{Tanabashi:2018oca,LHCb-PAPER-2014-035}.
    The predicted width for this $D_s^*$ state, interpreted as the $1^3D_1$ $s\bar{c}$ excitation in the quark model, is 197\mev~\cite{Godfrey:2015dva}.
    }
    \label{fig:bw2t1p6_159}
\end{figure}

The $4457\mev$ structure peaks near the $\Lc(2595)\Dzb$ threshold ($J^P=1/2^+$)~\cite{Liu:2015fea}. 
The best $h$ candidate for the corresponding triangle diagram is the $D_{s1}(2860)^-$ meson, which has a mass of $2859\pm27$\mev and a width of $159\pm80$\mev~\cite{Tanabashi:2018oca}.
The $P_c(4312)^+$ is not far from the $\Lc\Dstarzb$
threshold ($J^P=1/2^-$ or $3/2^-$).
Exchanging an excited $1^-$ $D_s^{**}$ meson with $M_0=3288$\mev produces the peak at 4312\mev in the narrow width approximation ($\Gamma_{0}\to0$). 
The $P_c(4440)^+$ is well above the $\chi_{c0}p$ threshold ($1/2^+$). 
Exchanging  an excited $1/2^+$ $\Lz^*$ with $M_0=2153$\mev produces 
a peak at the right mass when $\Gamma_0 \to 0$.
In fact, a good quality fit to all three $P_c^+$ peaks is obtained 
when $\Gamma_0$ is small, as illustrated in Fig.~\ref{fig:triangleFits} (top). 
However, this interpretation is unrealistic for the $P_c(4312)^+$ and $P_c(4440)^+$
peaks. When more plausible widths for the excited hadrons are used, such as 
$\Gamma_0=50$\mev,
no narrow peaks can be created above the thresholds, as shown in Fig.~\ref{fig:triangleFits} (bottom). The triangle-diagram hypothesis is more plausible for the $P_c(4457)^+$ state. 
An example fit using two BW terms and one triangle-diagram amplitude is shown in Fig.~\ref{fig:bw2t1p6_159}. The fit quality is lower than that obtained using three BW amplitudes. 
However, further investigation of this interpretation of the $P_c(4457)^+$ state is warranted within an amplitude analysis, which will provide greater discrimination between the triangle-diagram and BW amplitudes. 

\section{Comments on the results in Ref.~\cite{LHCb-PAPER-2015-029}}
\label{sec:argands}

The $P_c(4450)^+$ state reported in Ref.~\cite{LHCb-PAPER-2015-029} should be considered obsolete and replaced by the $P_c(4440)^+$ and $P_c(4457)^+$ states.
The six-dimensional amplitude analysis reported in Ref.~\cite{LHCb-PAPER-2015-029}, which provided evidence for the $P_c(4380)^+$ state, is also obsolete since it used the single $P_c(4450)^+$ state and it lacked the $P_c(4312)^+$ state. 
Therefore, the results presented in the Letter weaken  the previously reported evidence for the $P_c(4380)^+$ state, but do not contradict its existence, since the present one-dimensional analysis is not sensitive to wide $P_c^+$ states. 
Only a future six-dimensional amplitude analysis of $\Lb\to\jpsi p\Km$ decays that includes the $P_c(4440)^+$, $P_c(4457)^+$, and $P_c(4312)^+$  states will be able to determine if there is still evidence for the $P_c(4380)^+$ state or any other wide $P_c^+$ states. 

Reference~\cite{LHCb-PAPER-2015-029} performed  a cross-check of the six-dimensional amplitude model by replacing the BW function for the $P_c(4450)^+$ state, under the preferred quantum-number assignment, by complex amplitudes in six narrow $m_{\jpsi p}$ bins near the peak region. The complex amplitudes obtained from this cross-check had large statistical uncertainties. 
The systematic uncertainties were not evaluated.
This cross-check was repeated for the $P_c(4380)^+$ state. 
The results of both cross-checks were displayed as Argand diagrams, which were consistent with the expected phase motion from the BW functions (more so for the $P_c(4450)^+$ structure than for the $P_c(4380)^+$), but that should now be considered obsolete. The $P_c(4380)^+$ Argand diagram was obtained using a model in which the $P_c(4450)^+$ structure was represented by only one resonance and the $P_c(4312)^+$ state was not included. The $P_c(4450)^+$ Argand diagram was obtained using a model in which the $P_c(4312)^+$ state was missing, and under the assumption that only a single partial wave describes the structure peaking at 4450\mev. This assumption is difficult to justify given the two-peak observation presented in the Letter. Furthermore, since the natural widths of the $P_c(4440)^+$ and $P_c(4457)^+$ states are comparable to the mass resolution, the $P_c(4450)^+$ Argand diagram reported in Ref.~\cite{LHCb-PAPER-2015-029} would need to have the mass-resolution effects unfolded to probe the underlying complex phase motion even if both states happened to have the same quantum numbers.

\clearpage 
\newpage  

\addcontentsline{toc}{section}{References}
\bibliographystyle{LHCb}
\bibliography{main,standard,LHCb-PAPER,LHCb-CONF,LHCb-DP,LHCb-TDR}

\newpage
\centerline{\large\bf LHCb collaboration}
\begin{flushleft}
\small
R.~Aaij$^{28}$,
C.~Abell{\'a}n~Beteta$^{46}$,
B.~Adeva$^{43}$,
M.~Adinolfi$^{50}$,
C.A.~Aidala$^{80}$,
Z.~Ajaltouni$^{6}$,
S.~Akar$^{61}$,
P.~Albicocco$^{19}$,
J.~Albrecht$^{11}$,
F.~Alessio$^{44}$
% ...

{\footnotesize \it
% ...
$ ^{6}$Universit{\'e} Clermont Auvergne, CNRS/IN2P3, LPC, Clermont-Ferrand, France\\
% ...
$ ^{11}$Fakult{\"a}t Physik, Technische Universit{\"a}t Dortmund, Dortmund, Germany\\
% ...
$ ^{19}$INFN Laboratori Nazionali di Frascati, Frascati, Italy\\
% ...
$ ^{28}$Nikhef National Institute for Subatomic Physics, Amsterdam, Netherlands\\
% ...
$ ^{43}$Instituto Galego de F{\'\i}sica de Altas Enerx{\'\i}as (IGFAE), Universidade de Santiago de Compostela, Santiago de Compostela, Spain\\
$ ^{44}$European Organization for Nuclear Research (CERN), Geneva, Switzerland\\
% ...
$ ^{46}$Physik-Institut, Universit{\"a}t Z{\"u}rich, Z{\"u}rich, Switzerland\\
% ...
$ ^{50}$H.H. Wills Physics Laboratory, University of Bristol, Bristol, United Kingdom\\
% ...
$ ^{61}$University of Cincinnati, Cincinnati, OH, United States\\
% ...
$ ^{63}$Syracuse University, Syracuse, NY, United States\\
% ...
$ ^{80}$University of Michigan, Ann Arbor, United States, associated to $^{63}$\\
% ...
}
\end{flushleft}

\end{document}